# Developing an Operational Model for an Internet of Data–Based Trusted Data Space for Vocational Education

Yalan Dong, Master's degree, Beijing Philisense Technology Co., Ltd., Engineer (Intermediate Grade)
Hao Tian, PhD, Senior Engineer
Hongyang Jia, PhD, Engineer (Intermediate Grade)
Chunxue Wang, Master's degree, Project Management Professional (PMP)
Wenchang Lei, Master's degree

*Supported by National Key Laboratory of Data Space Technology and System*

## Abstract

Data circulation in vocational education has long been constrained by unclear ownership, a lack of trust, and insufficient incentives. The Internet of Data provides a new underlying technology for enabling trusted cross-domain data circulation. Building on an Internet of Data–based trusted data space for vocational education, this paper develops a multi-stakeholder operational model encompassing a value network, a product portfolio, revenue allocation, and governance rules. Emphasizing data sovereignty protection and incentive compatibility, the model provides a reference for the sustainable operation of regional vocational education data ecosystems.

## 1 Introduction

As emerging technologies such as artificial intelligence (AI) continue to advance, systems for supplying educational resources are gradually evolving to meet the needs of contemporary society. In 2025, the Central Committee of the Communist Party of China and the State Council issued the Outline for Building China into a Leading Country in Education (2024–2035), which calls for accelerating the development of a modern vocational education system, creating new forms of industry–education integration, and strengthening municipal industry–education consortia and industry-specific integration communities [1]. The research team behind the World Higher Education Digital Development Report identifies digital technology as a new productive force in educational development that drives systemic changes in teaching and learning. In particular, AI applications are gradually shifting the production of educational resources toward a model driven by human–machine collaboration, with governments, enterprises, educational institutions, and individuals jointly participating in the digital governance of educational resources [2].

Huang Ronghuai and colleagues argue that educational transformation involves data-based connections across time and space and the circulation of information. They also emphasize the importance of building evidence chains to assess the effects of AI-enabled comprehensive educational reform. Survey findings in the Report on the Digital Development of Vocational Education (2023 Edition) show that, on average, 88.92% of vocational institutions have achieved digital educational resource sharing across institutions (86.3%), between institutions and enterprises (86.77%), and within institutions (93.69%). However, only 33.42% of vocational institutions, on average, demonstrate a high level of resource sharing [3]. Consequently, most students at vocational institutions still lack opportunities to gain an in-depth understanding of actual workplace activities in enterprises after digitally supported workplace internships, indicating insufficient coordination between institutions and enterprises.

Improving the sharing of digital educational resources among vocational institutions and overcoming data silos and mutual-trust barriers have become important issues in vocational education reform. Trusted data spaces, proposed in the Action Plan for Trusted Data Space Development (2024–2028) released by the National Data Administration in 2024, can facilitate such sharing. A trusted data space is an infrastructure for data circulation and utilization that connects multiple stakeholders under consensual rules to enable the sharing and joint use of data resources. Many cities, enterprises, and industries have already established corresponding trusted data spaces. Their real purpose, however, lies in sustaining operations after construction, attracting additional participants, fostering endogenous motivation, and activating the circulation of data as a factor of production within the space, thereby promoting the development of the digital economy.

Focusing on a vocational education trusted data space underpinned by Internet of Data technology, this paper investigates an operational model that enables institutions, enterprises, students, and other stakeholders to benefit on a sustained basis.

## 2 Design Rationale for an Internet of Data–Driven Trusted Data Space for Vocational Education

### 2.1 Three Dilemmas in Vocational Education Data Circulation

During the digital transformation of vocational education, data circulation and sharing are essential to deepening industry–education integration and establishing a closed-loop approach to precision talent development. Yet data circulation in this sector has remained at a low level. The problem is manifested in three dilemmas: stakeholders are afraid to share, unwilling to share, or unable to sustain sharing. These reflect deeper tensions concerning ownership, security, and incentives.

First, the ownership dilemma makes stakeholders afraid to share. Vocational education data encompass students' personal information, institutional teaching-process records, and enterprises' internal job competency standards and production data, while the boundaries of data property rights remain unclear. Although the Personal Information Protection Law and the Data Security Law establish principles for classified and graded data protection, detailed rules have yet to clarify the allocation of rights to hold, use, and derive returns from data such as student skill profiles and practical-training behavior trajectories in educational contexts. To avoid disputes over infringement, institutions and enterprises tend to retain data within their internal systems [4][5].

Second, the security dilemma makes stakeholders unwilling to share. Educational data sharing mainly involves teaching data circulated within institutions. Competition among institutions, the absence of market-based mechanisms for teaching and research resources, and technology enterprises' commercial interests all weaken stakeholders' willingness to share. Furthermore, educational activities generate complex and sensitive data involving numerous stakeholders. Against the backdrop of open-source AI development, fully opening these data may create a range of ethical problems [6].

Finally, the incentive dilemma makes sharing unsustainable. Even when technology alleviates ownership and security concerns, data circulation remains difficult to sustain without a reasonable mechanism for returning benefits. Providers of high-quality data, particularly enterprises that possess industry hiring standards, lack direct and visible returns from one-way sharing, and their motivation declines over time. Most institution–enterprise cooperation remains project-based rather than deeply collaborative, and the parties lack clear and reasonable benefit-allocation mechanisms [7]. Sustainable industry–education integration therefore requires endogenous economic incentives.

Upgrading technical architectures alone is far from sufficient to resolve these dilemmas. An operational model is needed that clearly defines each party's rights and responsibilities and enables all parties to obtain corresponding benefits.

### 2.2 Integrating the Internet of Data with Trusted Data Spaces

Integrating the Internet of Data with trusted data spaces offers a feasible path that coordinates technological and institutional solutions. In essence, the integration establishes a secure and controllable virtual highway for data, equipped with finely specified rule-based tollgates, without changing where data are physically stored or who controls that storage. This allows data use rights to be separated from ownership.

The Internet of Data is a data interconnection network based on Digital Object Architecture (DOA) and a unified identifier resolution system. It assigns a globally unique identifier to each digital object, such as a set of academic records or a job competency model, enabling cross-system and cross-institutional discovery, resolution, addressing, and access. It thus addresses the basic interconnection questions of where data are located and how they can be found [8][9][10]. The Internet of Data does not itself store data; it indexes data locations and establishes access paths. This provides underlying network support for decentralized information exchange in which data remain within their originating domains.

A trusted data space is a governance framework that enables secure data sharing through use control rather than possession control under the principle of data sovereignty. Its core components include data connectors, usage-policy engines, and audit and evidence-preservation systems. Providers may specify data-use restrictions in advance, such as permitting aggregate statistical analysis only while prohibiting the identification of individuals. These policies are automatically enforced when data-access requests are made, and the entire process is recorded in tamper-proof logs [11][12].

The two approaches together form a complete closed loop: the Internet of Data addresses data discovery and interconnection across regions and institutions, while the trusted data space addresses data delivery and controlled use. In vocational education, an enterprise thousands of kilometers away can discover an institution's de-identified student skills data through the Internet of Data. Connectors and secure sandboxes within the trusted data space can then support agreed analyses, such as job competency matching, without exposing the underlying records. Access rights are revoked according to policy once computation is complete. This integration fundamentally addresses the ownership concerns underlying fear of sharing, because data remain local, and the security concerns underlying reluctance to share, through end-to-end use control and auditable records. It also creates the technical conditions for pay-per-use and value-based allocation incentives in the subsequent operational model.

## 3 Designing the Operational Model

### 3.1 Stakeholder Roles and Value Relationships

The vocational education trusted data space adopts a hybrid operating arrangement combining government guidance with an industry alliance. Education authorities are responsible for top-level planning, standard-setting, and compliance oversight, providing institutional endorsement and public credibility. The industry alliance, jointly formed by leading enterprises, exemplary vocational institutions, and third-party assessment organizations, handles daily operations, product iteration, and market development, maintaining an agile response to changing industrial needs. This arrangement is proposed because a purely government-led model can encounter administrative-efficiency bottlenecks and struggle to keep pace with rapid data-product iteration, whereas a purely market-driven model may lack public credibility and attract insufficient institutional participation. By assigning rule-making to government and operations to the alliance, the hybrid arrangement achieves a structural balance between authority and flexibility.

### 3.1.1 Role Composition and Functional Boundaries

Participants can be classified into six roles. These roles are not mutually exclusive: an institution may simultaneously act as both a data provider and a data user. Nevertheless, the boundaries of rights and responsibilities associated with each role must be clearly specified through digital contracts.

| Role | Typical participants | Core functions | Key rights and responsibilities |
|---|---|---|---|
| Data provider | Vocational institutions, industry–education integration enterprises, and vocational skills assessment organizations | Connect teaching-process data, job competency standards, and skills certification results to the space in standardized formats; publish data and grant authorization. | Demonstrate the lawful provenance and quality of data; retain the right to set usage restrictions and receive a share of revenue. |
| Data user | Employers, human resources service providers, and institutional teaching and research departments | Obtain required data products or services through discovery, transformation, and delivery to support recruitment decisions and curriculum optimization. | Use data only for agreed purposes; avoid retention beyond the authorized scope and secondary distribution; pay according to usage volume or subscription period. |
| Data service provider | (1) Data trading platforms and supply–demand matching providers;<br>(2) Internet of Data node operators and privacy-preserving computation providers;<br>(3) Value-added application developers and AI model training organizations. | (1) Match supply and demand, connect providers and users, and reduce transaction search costs;<br>(2) Provide managed operations, secure storage, trusted computation environments, and node maintenance;<br>(3) Undertake commissioned or joint development to transform raw data into products or analytical services. | (1) Do not withhold or alter supply–demand information; charge fees per successful match;<br>(2) Ensure availability and security under service-level agreements (SLAs); process only encrypted or de-identified data, without access to plaintext;<br>(3) Deliver outputs under development agreements; allocate intellectual property rights as contractually agreed. |
| Data subject | Enrolled students, graduates, and employed learners upgrading their skills | As rights holders of personal data, exercise rights to informed consent, withdrawal of | Determine the scope and duration of personal-data access through authorization agreements; |

| Role | Typical participants | Core functions | Key rights and responsibilities |
|---|---|---|---|
| | | authorization, and benefit sharing; a distinct role in vocational education. | receive benefit credits and a share of revenue. |
| Data space operator | Industry alliance comprising leading enterprises, exemplary institutions, and assessment organizations | Provide trust management and control; authenticate access for all relevant elements, apply dynamic controls throughout the process, and preserve traceable evidence across scenarios; manage daily operations and enforce rules. | Fulfil operational obligations under digital contracts; accept performance evaluation and compliance audits by the oversight authority. |
| Oversight authority for data-space development | Education authorities | Develop standards and rules, allocate dedicated funding, monitor operational conditions, evaluate performance, and implement corrective action. | Refrain from intervening in daily operations; exercise supervision over the operator and the power to revise rules. |

Students' personal data are often subsumed under the category of data providers. However, students are the central subjects of educational data [13], and protecting their privacy is a distinctive requirement in educational settings. Students are both the sources of data and the holders of the associated rights. Their rights to informed consent and benefit sharing must therefore be protected independently of institutions' data-provision activities, justifying a separate role.

### 3.1.2 Multi-Stakeholder Value Exchange

With the operator acting as the value hub, the trusted data space establishes a reciprocal value-exchange network linking the contributions and returns of all six roles throughout data onboarding, circulation, and value creation. Data providers contribute standardized teaching-process data, job competency standards, and skills certification results in exchange for data-product revenue shares and greater industry influence. Data users access trusted, processed products or analytical services to support recruitment and curriculum optimization and pay according to usage volume or subscription period. Data service providers contribute supply–demand matching, managed operations, and commissioned development throughout publication and delivery, in return for operations fees and development revenue shares. Data subjects—students—authorize the circulation of personal learning data in exchange for benefit credits redeemable for continuing education and professional certification, together with revenue sharing. The oversight authority supplies institutional arrangements and fiscal funding, providing the rules and financial support for the space in exchange for its contribution to vocational education governance objectives. In these exchanges, the operator both supplies trust management capabilities and executes value allocation. It offers access authentication, dynamic control, and evidence preservation and traceability, while distributing user payments under digital contracts among provider revenue shares, service-provider remuneration, data-subject benefit credits, and a data space development fund. Each data transaction is thus associated with clearly identifiable costs and returns. These relationships are illustrated below.

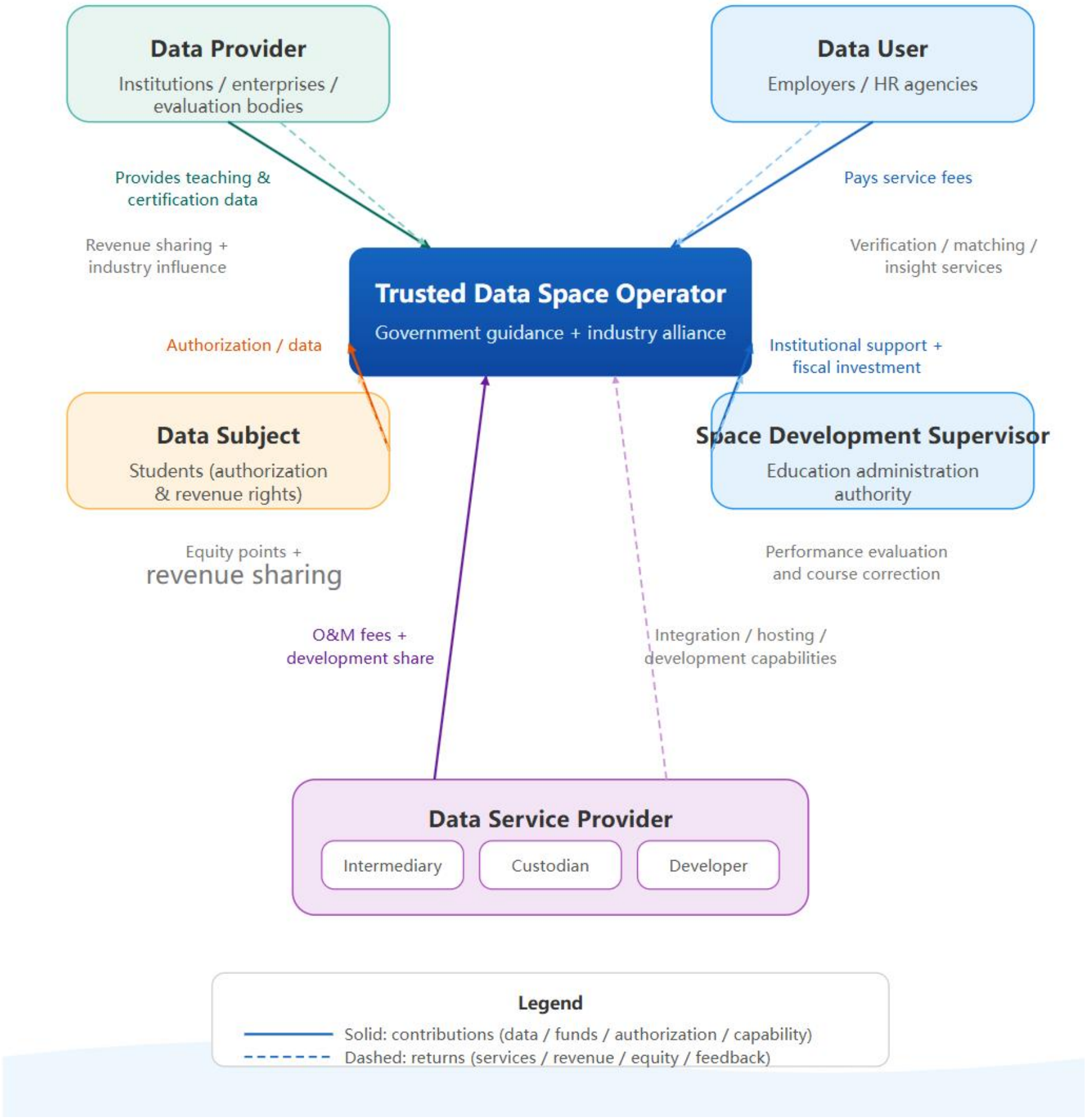

*Editable English schematic of the original diagram. Arrows indicate the direction of exchanges relative to the operator.*

Figure 3-1. Multi-stakeholder value exchanges in the vocational education trusted data space.

## 3.2 A Tiered Product Portfolio and Revenue Model

The product portfolio of the vocational education trusted data space consists of three tiers, corresponding respectively to public-service, market-oriented, and asset-related attributes.

The bottom tier comprises basic public services. These have the characteristics of public goods and are offered free of charge or funded by government expenditure to attract users with essential needs and establish trust. The middle tier comprises value-added analytical services, providing analytical capabilities for specific scenarios on a pay-per-use or subscription basis; these constitute the operator's principal revenue source. The top tier comprises data productization services, which process raw data in a compliant manner and package them into directly tradable data products. Revenue sharing at this tier unlocks the long-term economic value of data assets.

The three tiers progressively support one another. Users and data accumulated through basic services provide analytical samples and a market foundation for value-added services. Industry insights and modeling capabilities developed through value-added services, in turn, provide methodological support and quality assurance for data productization. This structure enables the data space to acquire users through public services at an early stage and sustain operations through high-value-added products in the medium and long term.

### 3.2.1 Basic Services Tier

The basic services tier serves as the public infrastructure of the data space. Its operating and maintenance costs are covered by dedicated education funding or participants' annual fees, and it is available free of charge to all registered users. Its core functions include the following:

(1) **Trusted verification of certificates and academic credits.** Employers and institutions admitting students to further study can verify the authenticity and validity of academic qualifications, vocational skill-level certificates, and credits online, replacing traditional paper-based verification. Integration with institutions' academic administration systems and assessment organizations' certification databases enables responses within seconds. Annual usage is projected to reach one million requests.

(2) **Public data queries.** A unified search entry point provides access to publicly available information, including program catalogs, curriculum standards, and industry talent-demand statistics, reducing information acquisition costs for institutional teaching and research departments and students.

(3) **Identity authentication and authorization management.** Students receive a view of their personal data assets and can review, grant, or revoke third-party access to their personal data at any time.

The value of basic services lies not in direct revenue generation but in creating indispensable use cases. When certificate verification becomes a standard recruitment step and credit inquiries become essential to admission reviews, user retention and network effects emerge naturally, providing a user-traffic foundation for higher-tier value-added services.

### 3.2.2 Value-Added Analytical Services Tier

Value-added analytical services target institutional users with a clear willingness and ability to pay. They are charged on a pay-per-use or annual subscription basis. Typical products include:

| Product | Target customers | Service content | Pricing model |
|---|---|---|---|
| Intelligent job–skills matching | Employers and human resources agencies | Compare enterprise job competency models with institutional training programs in a structured manner to generate candidate skills-match scores and gap-analysis reports. | Pay per request: individual requests or bulk packages. |
| Industry talent-demand insight reports | Vocational institutions | Integrate recruitment data from multiple sources and industry | Annual subscriptions tiered by report frequency and analytical depth. |

| Product | Target customers | Service content | Pricing model |
|---|---|---|---|
| | | developments to generate regional or sectoral skills-demand trends, forecasts of job increases and decreases, and recommendations for curriculum adjustments. | |
| Training-quality benchmarking | Vocational institutions | Benchmark graduate employment data and skills certification pass rates against peer institutions and industry standards across multiple dimensions to identify training deficiencies. | Annual subscription plus customized consulting. |
| Regional skills mapping | Local human resources and social security authorities and industrial parks | Use de-identified, aggregated data to map the regional distribution of workforce skills, supporting industrial policy and training-resource allocation decisions. | Project-based procurement. |

Value-added services follow a value-based pricing principle. Prices are not determined simply by data volume or computing resources, but by the recruitment costs saved, losses from misaligned training reduced, or decision-making efficiency improved through using the service. The data space can thereby capture the value added through data processing rather than function merely as a low-cost data conduit.

### 3.2.3 Data Productization Services Tier

The core logic of data productization services is to help data providers, particularly vocational institutions, package accumulated de-identified practical-training and teaching-process data into compliant products that meet circulation standards. These products are offered for a fee to third parties such as educational technology enterprises, curriculum development organizations, and purchasers of AI training data. The operator collects a service commission at an agreed rate.

The process has four key stages. First, during compliance review, the operator's compliance team conducts a personal information protection impact assessment of the proposed data products [14] and confirms that de-identification meets an irreversibility standard. Second, quality packaging involves format standardization, metadata annotation, and quality grading under industry data-product standards. Third, qualified products are listed in the data space marketplace, with usage boundaries and pricing strategies specified. Fourth, revenue is automatically allocated and settled according to usage records. Data providers receive the principal share of 60%–70%, the operator charges a service commission of 20%–30%, and the remainder is allocated to the data space development fund.

This tier primarily shifts the operator's role from service provider to trading-platform operator. Its income becomes linked to the market performance of data products, incentivizing continued improvement in product discovery, transaction matching, and compliance assurance.

### 3.3 Incentive-Compatible Allocation Mechanisms

The central goal of the allocation mechanism is incentive compatibility [15]: each participant, while maximizing its own interests, should also contribute to the growth of the data space's overall value. Three conditions must be satisfied. First, contributions must be measurable: inputs such as data, funding, and computing power must be quantifiable and recorded. Second, returns must be predictable: allocation rules must be transparent in advance so that participants can form stable expectations. Third, breaches must carry consequences: free-riding and opportunistic behavior must result in meaningful penalties.

#### 3.3.1 Contribution–Reward Mechanisms

Contribution–reward mechanisms aim to reduce data providers' perceived participation costs, turning data contribution from a one-way expenditure into a reciprocal activity that produces returns. The proposed arrangements are as follows:

(1) **Enterprises: exchange updated job requirements for services.** Enterprises submit or update structured job competency standards, including skills requirements, experience thresholds, and certification preferences. Each valid update earns a quota of free job–skills matching requests. Updating job standards contributes high-value industry knowledge to the data space, while free matching requests provide precisely the service enterprises need. Greater contributions generate more service benefits, creating a positive cycle. Frequent updates also keep job models current, benefiting all users.

(2) **Students: earn benefit credits by authorizing data use.** Students authorize the use of personal learning data, including course grades, practical-training records, and skills certifications, within the data space. The system automatically accumulates benefit credits based on data type and authorization duration. Credits can be redeemed for three types of benefits: discounts on continuing education courses at partner institutions, fee reductions or waivers for vocational skills certification examinations, and access to individual versions of value-added reports. The credit system follows a principle of small, frequent rewards accumulated over time, avoiding short-term behavior induced by large one-off incentives and encouraging sustained authorization. Students may withdraw authorization at any time. Credits already earned are not reclaimed, but no further credits accrue after withdrawal, balancing incentive effectiveness with protection of data-subject rights.

(3) **Institutions: exchange data-contribution rankings for resources.** Institutional contributions—measured by composite indicators including data onboarding volume, update frequency, and quality scores—are incorporated into annual evaluations and linked to the weighting used in allocating dedicated education funding. High-ranking institutions receive preferential treatment in the following year's infrastructure subsidies, value-added service fee reductions, and priority access to joint research projects. By linking data contribution to institutions' core interests in fiscal resources and academic reputation, this mechanism overcomes data inertia.

#### 3.3.2 Cost-Sharing Mechanisms

The data space's cost structure comprises infrastructure construction costs and ongoing operating costs, with different sharing strategies applied to each:

| Cost category | Main components | Sharing arrangement | Responsible parties |
| --- | --- | --- | --- |
| Infrastructure | Internet of Data node | Dedicated education | Education authorities as |

| Cost category | Main components | Sharing arrangement | Responsible parties |
|---|---|---|---|
| construction | deployment, privacy-preserving computation platform construction, and security authentication systems. | funding plus one-off alliance joining fees. | the principal funder, with initial alliance members sharing costs. |
| Routine operations and maintenance | Node maintenance, security monitoring, and basic-service operations. | Annual membership fees tiered by organization type and size. | All alliance members. |
| Value-added service development | Analytical model research and development, product iteration, and marketing. | An R&D reserve of 15%–20% of value-added revenue. | Funded from value-added revenue. |
| Data space development fund | Reserves for technology upgrades, incubation of new use cases, and risk buffering. | Reserve 10% of value-added revenue and data productization proceeds. | Funded from total operating revenue. |

### 3.3.3 Revenue Allocation Rules

Value-added revenue and data productization proceeds are allocated using a two-factor weighted model combining data contribution and operational services. Operating costs and development-fund reserves are first deducted from each distributable revenue amount. The remainder is allocated between providers and the operator according to the following formulas:

Data provider's return = Distributable revenue × Data contribution weight (recommended weight: 0.6–0.7);

Operator's return = Distributable revenue × Operational service weight (recommended weight: 0.3–0.4).

The data contribution weight is dynamically adjusted according to data scarcity, timeliness, and quality scores. The operational service weight covers the combined contributions of platform maintenance, compliance review, market matching, and technical support [16].

When a product incorporates data from multiple providers, the providers' share is further subdivided in proportion to their respective data contributions. Contribution is measured along dimensions such as the number of records, field completeness, timeliness of updates, and compliance rates for de-identification. The system automatically collects these metrics and generates contribution reports, reducing disputes arising from subjective judgments.

## 3.4 Trusted Governance Rules

### 3.4.1 Identity Onboarding and Authentication

The data space adopts decentralized identity (DID) as its unified identity onboarding approach. When a participant first joins, an authentication node authorized by the operator issues a DID identifier. Identity metadata, including organization type, qualification documents, and authorization scope, are recorded on a distributed ledger in the form of verifiable credentials (VCs) [17]. All subsequent data interactions use DIDs for authentication, enabling one-time authentication and mutual recognition throughout the space.

Identity onboarding comprises three steps. First, participants submit the relevant qualification documents; an authentication node reviews them and issues a DID upon approval. Second, participants sign digital contracts with the operator specifying the scope of data use, liability for breach, dispute resolution mechanisms, and other terms. These contracts are deployed as smart contracts, with key provisions executed automatically. Third, participants receive initial permissions specifying the data they may access, the services they may invoke, and the restriction policies they may establish.

Compared with conventional centralized account systems, the DID approach has two advantages: (1) identity data are not monopolized by a single platform, allowing participants to reuse the same identifier across different data spaces and application scenarios; and (2) cryptographic authentication does not require disclosure of underlying sensitive information. For example, students can prove the authenticity of academic qualifications without showing their national identification numbers to enterprises.

### 3.4.2 Data Usage Control

Data usage control is central to trusted governance. It enables providers to retain effective constraints on how their data are used after sharing, rather than losing control once data are handed over. The data space achieves this through three complementary mechanisms: advance policy specification, automated enforcement, and end-to-end auditing.

(1) **Advance policy specification: providers independently define restrictions.** At onboarding, providers use structured policy templates to specify usage constraints. Four common policy types are supported. Purpose restrictions include “aggregate statistical analysis only; individual-level queries prohibited” and “use limited to this recruitment project; authorization expires automatically when the project ends.” Operation restrictions include “reading permitted; downloading prohibited” and “model training permitted; display of source text prohibited.” Redistribution restrictions include “forwarding to third parties prohibited” and “internal use within this organization only.” Time restrictions include “authorization valid for 12 months; access rights automatically revoked upon expiry.”

(2) **Automated enforcement: the data space mandates policy compliance.** Usage policies are embedded in the data-access control layer, and the policy enforcement engine automatically checks every request. Requests that violate policy constraints are rejected and logged, without relying on users' voluntary compliance. For scenarios requiring privacy-preserving computation, such as joint modeling, enforcement extends into the computation environment. Computation is performed within a trusted execution environment (TEE) or a federated learning framework, and consumers receive only the results without access to raw data at any stage.

(3) **End-to-end auditing: tamper-proof operation records.** All data access, computation, and transmission operations generate audit logs containing the acting participant's DID, timestamp, operation type, scope of data involved, and policy-validation results. Logs are stored as hash chains to prevent subsequent tampering. Providers can inspect usage records for their data at any time; the operator regularly produces compliance audit reports; and regulators may retrieve audit chains for enforcement investigations under legally prescribed circumstances.

### 3.4.3 Maintaining Trust and Addressing Violations

The long-term operation of the data space depends on trust among participants. The trust-maintenance mechanism links reputation scores to functional permissions, translating an abstract trust relationship into an actionable governance instrument.

Each participant receives an initial reputation score, for example 100 points, which is dynamically adjusted according to its behavioral record. Sustained high-quality data contributions, proactive reporting of data-quality issues, and active participation in standard-setting can increase the score. Using data beyond the authorized scope, failing to update data on time and thereby disrupting downstream services, or refusing to cooperate with compliance audits can reduce it. Score changes are triggered automatically by the system, while major deductions require review and confirmation by the operator.

The relationship between reputation scores and permissions is specified below:

| Reputation score | Permission status | Measures |
|---|---|---|
| 90–100 | Full permissions | Access all authorized data; participate in data productization; enjoy priority service channels. |
| 70–89 | Standard permissions | Retain normal access to authorized data; suspend eligibility for data productization; withdraw priority services. |
| 50–69 | Restricted permissions | Limit access to basic services; suspend value-added service requests; require a corrective action plan. |
| Below 50 | Suspension or expulsion | Freeze all data-access permissions and initiate a breach investigation. In serious cases, expel the participant and enforce liability under the digital contract. |

This mechanism creates a closed loop in which behavior affects reputation and reputation determines permissions. To maintain or improve their permission levels, participants have incentives to comply with rules and maintain data quality, while violations continue to affect future benefits. A pathway for reputation recovery remains available: after completing corrective action and passing a review, restricted participants can gradually restore their scores, avoiding the excessive penalty of permanent exclusion for a single mistake.

### 3.4.4 Evolution of Governance Rules

Governance rules are not static; they are continually updated as technologies evolve, business activities expand, and regulatory conditions change. Rule revision follows five stages: proposal, deliberation, voting, pilot implementation, and formal adoption. Any alliance member may submit an amendment proposal. The operator organizes an open consultation involving relevant stakeholders for at least 30 days. A governance committee, composed of government, institutional, and enterprise representatives and technical experts in fixed proportions, votes on the proposal. Approved amendments are piloted on a limited scale for one quarter and formally adopted after their effects are evaluated. This process balances stability and adaptability, preventing uncertainty caused by frequent changes while avoiding governance failures caused by rigid rules.

## 4 Conclusions and Outlook

From the perspective of operational-model design, this paper develops a four-dimensional framework for a vocational education trusted data space, integrating stakeholder roles, tiered products, incentive allocation, and trusted governance. Whether a data space can operate sustainably depends not on its technical architecture but on whether it establishes a mechanism of commensurate rights and responsibilities and reciprocal benefits, shifting data circulation from policy-driven to market-driven operation. Future research may proceed in three directions: calibrating key incentive parameters through multi-agent simulation; conducting two to three years of empirical performance tracking through provincial pilot projects; and exploring identity mutual recognition and revenue coordination across data spaces.